\documentclass[12pt]{article}
\usepackage{graphicx}

\newcommand\pubnumber{SNSN-323-63}
\newcommand\pubdate{\today}

\def\kyoto{Department of Physics, Graduate School of Science, Kyoto University,Kitashirakawa Oiwake-cho, Sakyo-ku, Kyoto 606-8502, Japan}
\def\jaxa{Institute of Space and Astronautical Science (ISAS)/JAXA,
3-1-1 Yoshinodai, Chuo-ku, Sagamihara, Kanagawa 252-5210, Japan}
\def\kek{Institute of Particle and Nuclear Studies, High Energy Accelerator Research Org., KEK, 1-1 Oho, Tsukuba 305-0801, Japan}
\def\miya{Department of Applied Physics, Faculty of Engineering, University of Miyazaki,1-1 Gakuen Kibana-dai Nishi, Miyazaki 889-2192, Japan}
\def\rika{Department of Physics, Faculty of Science and Technology, Tokyo University of Science, 2641 Yamazaki, Noda, Chiba 278-8510, Japan}

\def\Title#1{\begin{center} {\Large #1 } \end{center}}
\def\Author#1{\begin{center}{ \sc #1} \end{center}}
\def\Address#1{\begin{center}{ \it #1} \end{center}}

\newcommand\pubblock{\rightline{\begin{tabular}{l} \pubnumber\\
         \pubdate  \end{tabular}}}
\newenvironment{Abstract}{\begin{quotation}  }{\end{quotation}}
\newenvironment{Presented}{\begin{quotation} \begin{center} 
             PRESENTED AT\end{center}\bigskip 
      \begin{center}\begin{large}}{\end{large}\end{center} \end{quotation}}

\def\beq{\begin{equation}}
\def\eeq#1{\label{#1}\end{equation}}
\def\eeqn{\end{equation}}

\def\beqa{\begin{eqnarray}}
\def\eeqa#1{\label{#1}\end{eqnarray}}
\def\eeqan{\end{eqnarray}}

\let\bar=\overbar

\def\Dslash{\not{\hbox{\kern-4pt $D$}}}
\def\dslash{\not{\hbox{\kern-2pt $\del$}}}

\def\msb{{\bar{\ssstyle M \kern -1pt S}}}

\begin{document}
\begin{titlepage}
\pubblock

\vfill
\Title{Investigation of the Kyoto's X-ray Astronomical SOIPIXs with Double-SOI Wafer for Reduction of Cross-talks}
\vfill
\Author{ Shunichi Ohmura$^{1}$, Takeshi Go Tsuru$^{1}$, Takaaki Tanaka$^{1}$, Ayaki Takeda$^{1}$, Hideaki Matsumura$^{1}$, Ito Makoto$^{1}$, Shinya Nakashima$^{2}$, Yasuo Arai$^{3}$, Koji Mori$^{4}$, Ryota Takenaka$^{4}$, Yusuke Nishioka$^{4}$, Takayoshi Kohmura$^{5}$, Kouki Tamasawa$^{5}$}
\Address{$^{1}$\kyoto \\ $^{2}$\jaxa \\ $^{3}$\kek \\ $^{4}$\miya \\ $^{5}$\rika}
\vfill
\begin{Abstract}
We have been developing X-ray SOIPIXs, "XRPIX", for future X-ray astronomy satellites. 
XRPIX is equipped with a function of  "event-driven readout", which allows us to readout signal hit pixels only and realizes a high time resolution ($\sim10\mu{\rm s}$). 
The current version of XRPIX suffers a problem that the readout noise in the event-driven readout mode is higher than that in the the frame readout mode, in which all the pixels are read out serially. 
Previous studies have clarified that the problem is caused by the cross-talks between buried P-wells (BPW) in the sensor layer and in-pixel circuits in the circuit layer. 
Thus, we developed new XRPIX having a Double SOI wafer (DSOI), which has an additional silicon layer (middle silicon) working as an electrical shield between the BPW and the in-pixel circuits. 
After adjusting the voltage applied  to the middle silicon, we confirmed the reduction of the cross-talk by observing the analog waveform of the pixel circuit. 
We also successfully detected $^{241}$Am X-rays with XRPIX. 

\end{Abstract}
\vfill
\begin{Presented}
International Workshop on SOI Pixel Detector (SOIPIX2015) \\
Tohoku University, Sendai, Japan, 3-6, June, 2015
\end{Presented}
\vfill
\end{titlepage}
\def\thefootnote{\fnsymbol{footnote}}
\setcounter{footnote}{0}

\section{Introduction}

The X-ray charge-coupled device (CCD) is the standard imaging spectrometer for the latest X-ray astronomy satellites. 
It offers the Fano limited spectroscopy with the energy resolution reaching $\sim$130 eV in FWHM at 6 keV and fine imaging with the small pixel size of $\sim$20-30$\mu$m and with the large imaging area of $20$-$30{\rm mm}$. 
However, the time resolution is poor ($\sim1$-$10{\rm sec}$), which limits fast timing observation of compact objects, such as black holes and neutron stars. 
Thus, we have been developing X-ray silicon-on-insulator pixels (SOIPIXs), "XRPIX", for future X-ray astronomy satellites. 

SOIPIX consists of three layers: The CMOS circuit layer made with low-resistivity silicon, a high-resistivity depleted silicon layer for X-ray detection, and a buried oxide layer (BOX layer) for insulation between the two layers. 
Each pixel has a sense-node of p$^{+}$ implanted in a sensor layer and a contact via implimented in the BOX layer which connects the p$^{+}$ with the CMOS circuit. A buried p-well (BPW) is implanted around the p$^{+}$ to collect signal charge and to suppress the back-gate effect of the CMOS circuit. 
XRPIX contains a comparator circuit in each pixel, and outputs event hit trigger timing signal and the row and column addresses of the hit pixel. 
It enables the function of "event-driven readout", with which we can readout hit pixels only and hence realize a high time resolution reaching $\sim10\mu{\rm sec}$. 

Takeda et al.(2014) reported that the spectral performances of the current version of XRPIX in the event-driven readout mode is significantly lower from those in frame-readout mode, in which all the pixels are read out serially like a CCD\cite{TakedaPoS2014}. 
The authors suggests that one of the causes is capacitive coupling between the trigger signal in the CMOS circuit layer 
and the sense-node and BPW in the sensor layer. 
Thus, we adopt a Double-SOI wafer (DSOI)\cite{MiyoshiNIMA2013, HondaPoS2014, HaraVertex2014}. 
The newly introduced middle silicon layer between the circuit and sensor layers in a DSOI wafer is expected to act as an electrostatic shield and  to reduce the capacitive coupling between them. 

In this paper, we report first experimental results from the Double SOI version of XRPIX3 (XRPIX3-DSOI)\cite{TakedaIEEE-NSS2013}. 
The middle silicon layer is processed in the whole chip region except contact via connecting the sense-node and circuit layer (Figure 1). 
The thickness of the middle silicon layer is 80nm. 
Both of  the distances between the middle silicon layer and circuit layer and between the middle silicon layer and sensor layer are 160nm.
XRPIX3-DSOI has a pixel size of 30$\times$30$\mu$m$^{2}$, a BPW size of 14$\times$14$\mu$m$^{2}$. 
It has a format of 32$\times$32 pixels, where there are two different types of test element group (TEG) pixels. 
A half of the 32$\times$32 pixels has charge-sensitive amplifier (CSA) and the other half has source follower (SF). 
All the results shown in the paper were obtained with the TEG of SFs operated in the frame-readout mode. 

\begin{figure}[htb]
\centering
\includegraphics[height=3in]{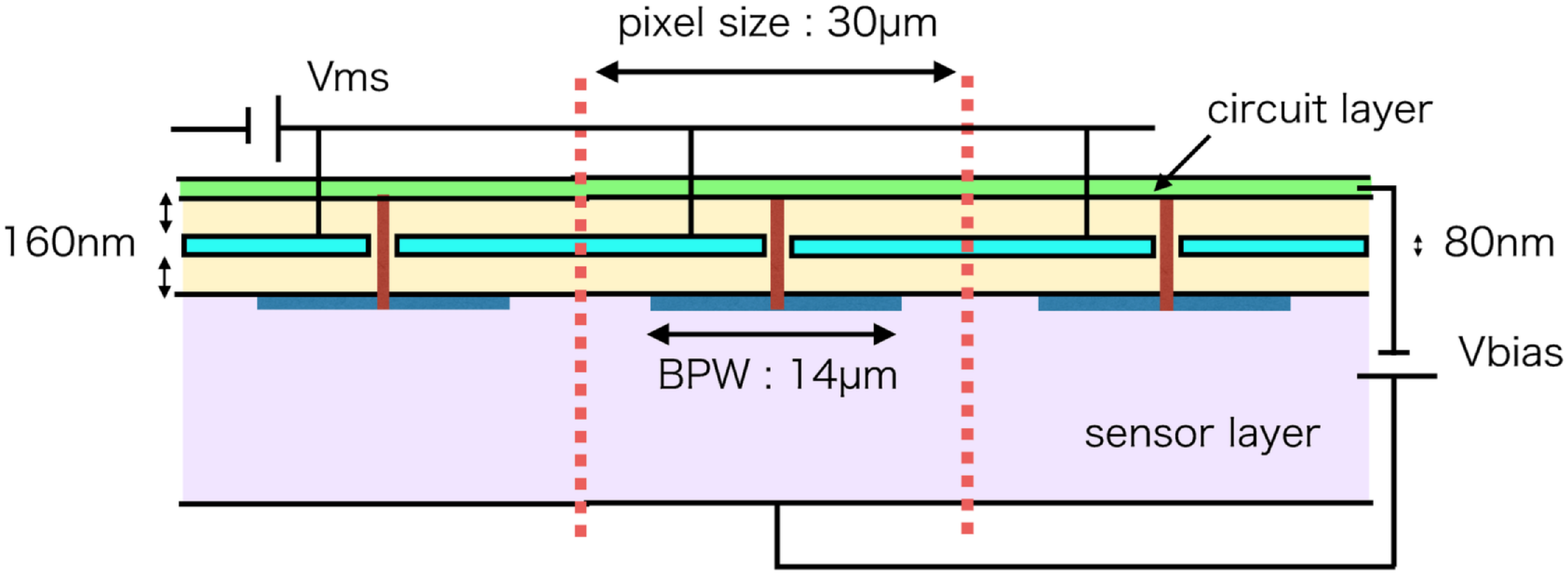}
\caption{The cross-sectional view of the XRPIX3-DSOI}
\label{fig:magnet}
\end{figure}

\section{Experimetal results}

\subsection{Laser irradiation test on XRPIX3-DSOI}
We irradiated XRPIX3-DSOI with laser light at room temperature (25$^\circ{\rm C}$) as first experiment. 
We observed the images changing the voltage of the middle silicon layer (V$_{ms}$) from -1V to +1V while the back bias voltage (V$_{bias}$) was fixed at 30V.
Figure~2 shows the laser light images we obtained in the frame-readout mode with the exposure time of $1{\rm msec}$. 
The laser light is detected when the middle silicons voltage is above +0.6V and its pulse height becomes higher as the voltage is increased. 
This result can be explained as follows. 
The voltage of sense-node and BPW is 0V and the back bias voltage is positive, 
which means that the positive voltage is required for the electric field in sensor layer to converge into the BPW and sense-node as shown in Figure~3.  
The charge-collection efficiency increases as a higher voltage is applied to the middle-silicon layer. 

\begin{figure}[htb]
\centering
\includegraphics[height=1.5in]{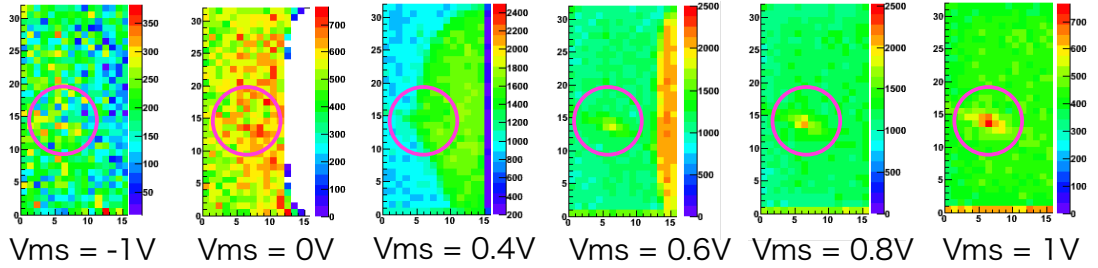}
\caption{The laser light images of XRPIX3-DSOI with the middle silicon voltages from -1V to +1V. 
The magenta circle indicates the position where the laser light is illuminated. 
}
\label{fig:magnet}
\end{figure}

\begin{figure}[htb]
\centering
\includegraphics[height=1in]{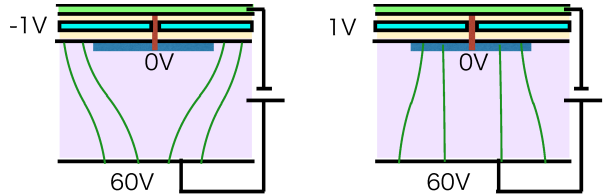}
\caption{Left and right panels show the schematic views of the electric fields in sensor layer when the voltage applied to the middle silicon layer is -1V  and +1V, respectively. }
\label{fig:magnet}
\end{figure}

\subsection{Cross-talk between the sensor and circuit layers}
In order to see if DSOI reduces the cross-talk, we observed the analog waveforms of the pixel circuits of XRPIX3-DSOI and single SOI version of XRPIX3b (XRPIX3b-SSOI). 
XRPIX3b has the same pixel circuit as XRPIX3 and hence we can make direct comparison with each other. 
The left panel of Figure~4 shows the waveform of XRPIX3b-SSOI when we irradiate with laser light. 
The rise of the analog signal before the illumination of the laser light is due to dark current. 
The laser light illumination starts at the timing of (1), which follows by the further rise of analog signal. 
The positive edge trigger is asserted at the timing of (2) when the analog signal exceeds the threshold. 
A jump of $\sim$80mV is seen in the analog signal at the timing of (2), 
which is the cross-talk between the sense-node and the trigger signal as is reported by Takeda et al.(2014)\cite{TakedaPoS2014}. 
On the other hand, XRPIX3-DSOI shows no jump in its analog waveform at the timing of trigger out (the right panel of Figure~4). 
This result suggests that we successful reduce the cross-talks by introducing Double-SOI.

\begin{figure}[htb]
\centering
\includegraphics[height=6cm]{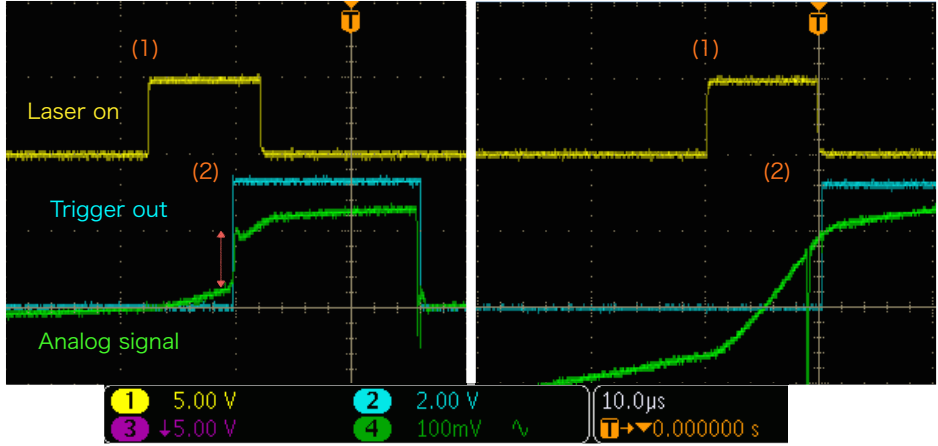}
\caption{The analog waveforms of the pixel circuits of XRPIX3b-SSOI (left panel) and XRPIX3-DSOI (right panel) when laser light is illuminated.
}
\label{fig:magnet}
\end{figure}

\subsection{X-ray Spectrum with XRPIX3-DSOI}

\begin{figure}[htb]
\centering
\includegraphics[height=2in]{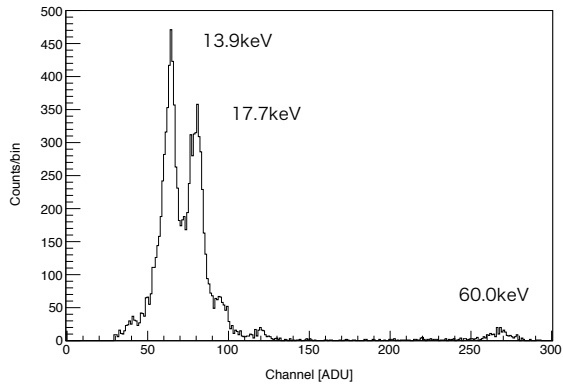}
\caption{the spectra of $^{241}$Am X-ray with XRPIX3-DSOI}
\label{fig:magnet}
\end{figure}
XRPIX3b-SSOI has the same in-pixel analog circuit and layout, and the same on-chip readout analog circuits as XRPIX3-DSOI. 
Having the middle silicon layer or not is the only difference between the two devices concerning the gain. 
The gain is determined mainly by the parasitic capacitance at sense-node\cite{TakedaJINST2015}. 
In the case of XRPIX3b-SSOI, the capacitance between the circuit layer and the BPW electrically connecting to the sense-node dominates the parasitic capacitance . 
On the other hand, the dominant capacitance in the case of XRPIX3-DSOI is the one between the middle silicon and the BPW. 
The ratio of the dominant capacitances of the two devices is 0.8, which is calculated from the ratio of the thicknesses of the insulators (160nm for XRPIX3-DSOI and 200nm for XRPIX3b-SSOI). 
Thus, we expect XRPIX3-DSOI has the 0.8 times lower gain than XRPIX3b-SSOI. 

Figure~5 shows the X-ray spectrum of $^{241}$Am obtained with XRPIX3-DSOI in the frame-readout mode. 
The device temperature, back bias voltage and middle silicon voltage are $-40^\circ{\rm C}$, +60V and +1V, respectively. 
We note that this is the first X-ray spectrum obtained with the Double SOI type of SOIPIX sensors. 
The energy calibration shows the output gain of 4.4 $\mu{\rm V/e}^-$. 
It is 0.8 times lower than the gain of XRPIX3b-SSOI (5.4 $\mu{\rm V/e}^-$). 
The result agrees with the expectation. 

Takeda et~al. (2015) shows that the readout noise of the XRPIX series (from XRPIX1 to XRPIX3b) is inversely proportional to the gain\cite{TakedaJINST2015}. 
According to Figure~5 of Takeda et~al. (2015), we expected the readout noise of XRPIX3-DSOI to be 103 ${\rm e}^-$(rms). 
However, the actual readout noise of XRPIX3-DSOI obtained with the pedestal peak was 125 ${\rm e}^-$(rms). 
It is significantly igher than the expectation. 
Although the cause of that is still unclear, we suspect the electric noises in the voltage applied to the middle silicon layer. 
We will do further investigation in future. 

\section{summary}
We have been developing XRPIX using the SOI CMOS technology for future X-ray astronomy satellites. 
We introduced a Double-SOI wafer and developed XRPIX3-DSOI for the middle silicon to reduce the cross-talk between the sensor and the circuit layers. 
We revealed that applying the middle silicon at a voltage higher than +0.6V is necessary to collect signal charge and obtain an image. 
We confirmed the introduction of DSOI reduces the cross-talk by observing the analog waveform as expected. 
We successfully obtained X-ray spectrum of of $^{231}$Am. 
The gain of XRPIX3-DSOI is 0.8 times lower than that of XRPIX3-SSOI, 
which is explained by the parasitic capacitance between the BPW and the middle silicon layer.


\end{document}